\documentclass[twocolumn,showpacs,preprintnumbers,amsmath,amssymb,prl,floatfix]{revtex4}

\usepackage{graphicx}
\usepackage{dcolumn}
\usepackage{bm}
\newcommand{\bra}[1]{\langle #1 |}
\newcommand{\ket}[1]{| #1 \rangle}
\newcommand{\braket}[2]{\langle #1 | #2 \rangle}

\newcommand{\creation}[1]{\hat{#1}^{\dag}}
\begin{document}

\title{Collisional decay of a strongly driven Bose-Einstein condensate}
\author{N. Katz}
\author{E. Rowen}
\author{R. Ozeri\footnote{Current address: Time and Frequency Division
NIST 325 Broadway Boulder, Colorado 80305, USA}}
\author{ N. Davidson}
\affiliation{Department of Physics of Complex Systems,\\
Weizmann Institute of Science, Rehovot 76100, Israel}
\date{\today}

\begin{abstract}
We study the collisional decay of a strongly driven Bose-Einstein
condensate oscillating between two momentum modes. The resulting
products of the decay are found to strongly deviate from the usual
s-wave halo. Using a stochastically seeded classical field method
we simulate the collisional manifold. These results are also
explained by a model of colliding Bloch states.
\end{abstract}


\maketitle

The decay of many-body states coupled to a quasi-continuum of
collisional products is a topic of great experimental and
theoretical interest \cite{enhance-supp,RMP}. Experimental
Bose-Einstein condensation (BEC) allows us to investigate this
decay in detail by use of highly controlled optical lattice
potentials. Both the coherent evolution of the condensate in the
lattice, and the nature of the quasi-continuum can be manipulated
and quantified \cite{igor}.

The finite lifetime of perturbative bulk excitations in BEC,
namely Beliaev and Landau damping, has been extensively studied
\cite{beliaev,ours-beliaev,beliaev-foot,jin} and is rather well
understood. These studies were extended recently to the ground
state of a BEC in an optical lattice, and weak excitations over
such a state, using band theory formulation
\cite{theory-lattice,observation-inguscio,arimondo-BO}.

Coherent Rabi oscillations of the condensate between two (or more)
macroscopically-populated momentum states can be driven by a
strong moving optical lattice potential. These oscillations are
described as beating between Bloch states belonging to different
bands of the lattice \cite{phillips-lattice,ours-dephasing}. Due
to interaction nonlinearity, a superposition of two such beating
Bloch states no longer solves the time-dependent Schr\"{o}dinger
equation, and thus the expected spectra and dynamics are richer
than for a single Bloch state \cite{stamper-kurn}. The decay of
these excited states cannot be described by mean-field theory nor
as an interaction between perturbative Bogoliubov quasi-particles
\cite{ours-3wm} as in the Beliaev formalism.

In this letter we study the collisional decay of such a strongly
driven BEC undergoing coherent Rabi oscillations between momentum
states by a resonant two-photon Bragg process
\cite{ketterle-bragg}. We measure a clear deviation of the
collisional products from the s-wave halo observed for collisions
of a weak excitation with the BEC \cite{enhance-supp}. Using a
stochastically seeded classical field Gross-Pitaevkii equation
(GPE) simulation \cite{gardiner}, we observe similar decay
dynamics. These results are then explained by a model which
includes collisions between Bloch states of the optical lattice as
a perturbation.

As in \cite{ours-bragg}, our nearly pure ($\sim 95\%$) BEC of
$N=1.6 (\pm 0.5)\times 10^{5}$ $^{87}$Rb atoms in the
$|F,m_{f}\rangle =|2,2\rangle $ ground state, is formed in a
magnetic trap with radial and axial trapping frequencies of
$\omega_r=2\pi \times 226$ Hz and $\omega_z=2\pi \times 26.5$ Hz,
respectively, leading to a healing length $\xi=0.23$ $\mu$m. The
condensate is driven by a pair of strong Bragg beams
counter-propagating along the axial direction $\hat{z}$, with
wavenumbers $k_{d1}=-k_L \text{ and }k_{d2}=k_L$, with
$k_L=2\pi/780$ nm. The laser frequency is red-detuned 44 GHz from
the $^{87}$Rb D$_2$ transition in order to avoid spontaneous
emission. The depth of the resulting optical lattice potential is
characterized by the two-photon Rabi frequency $\Omega_d$. For
strong excitations the mean-field shift is largely suppressed
\cite{ours-dephasing}, due to temporal averaging of the shift to
zero during a cycle of oscillation, and hence the frequency
difference between the dressing beams (in the laboratory frame) is
set to $\delta_d=2\pi\times15$ kHz, the free-particle resonance.
This leads to Rabi-like oscillations between the momentum states
$k=0$ and $k=2k_L$.

\begin{figure}[h]
\begin{center}
\includegraphics[width=7cm]{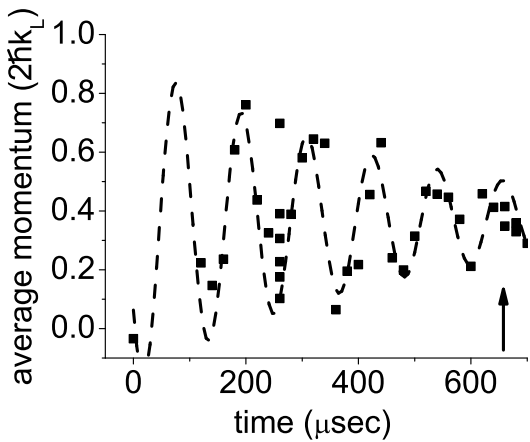}
\end{center}
\caption{ \label{fig:oscillations} Average momentum per particle
contained in the atomic cloud as a function of time (in units of
$2\hbar k_L$, in the laboratory reference frame). Oscillations are
due to a strong moving optical lattice which is suddenly switched
on, leading to Rabi oscillations between momentum wavepackets. The
decay in the oscillations (fit by the dashed line) is mainly due
to collisions which deplete the condensate. The arrow marks the
point at which Fig. \ref{fig:tomog}(a) (and the subsequent
theoretical figures) are taken.}
\end{figure}

The oscillation in the momentum of the atoms is apparent in Fig.
\ref{fig:oscillations}, where we plot the measured average
momentum per particle in the $\hat{z}$ direction as a function of
time, extracted from time of flight images. The oscillation
frequency, as obtained by a decaying sinusoidal fit, is
$\Omega_d/2\pi=8.6$ kHz.

In the strongly driven condensate, both finite size broadening,
and inhomogeneous density broadening are greatly suppressed
\cite{ours-dephasing}. Therefore the decay of the oscillations is
mostly due to the collisions between atoms in momentum modes 0 and
$2\hbar k_L$. The products of such a collision have an average
momentum of $\hbar k_L$, and do not, in general, oscillate any
more in momentum space. For a Bogoliubov excitation, which is a
weak excitation of momentum $2\hbar k_L$ over a large condensate
of zero momentum, the collisional products are known to be located
on a  shell in momentum space, known as the s-wave halo. This
shell is the surface in momentum space, conserving both energy and
momentum for the collision. Due to the Bogoliubov dispersion
relation, which is quadratic for $2k_L\xi>1$, this shell is nearly
spherical.

\begin{figure}[h]
\begin{center}
\includegraphics[width=7cm]{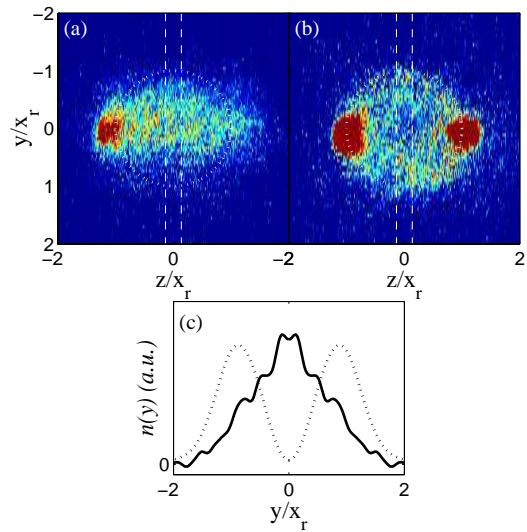}
\end{center}
\caption{ \label{fig:tomog} (a,b) Absorption images after
$t_{tof}=38$ ms time of flight following a resonant dressing
pulse. (a) Strong pulse $\Omega_d/2\pi=8.6$ kHz, pulse width
$t=660$ $\mu$s (b) Weak pulse $\Omega_d/2\pi<2$ kHz, pulse width
$t=370$ $\mu$s. Dotted circles  represent the predicted s-wave
shell. $x_r=\hbar k_L/M\times t_{tof}$ is the ballistic expansion
distance of an atom with wavenumber $k_L$ (in lattice frame of
reference). The collisional manifold for the strong pulse is
clearly shifted inwards as compared to that of the weak pulse,
which agrees with the expected s-wave shell. (c) The density
distribution along the $y$-axis obtained by computerized
tomography \cite{ours-tomography} of the data in (a,b) averaged
over a slice marked by the vertical dashed lines. The solid line
is for the strongly driven BEC of (a), and the dotted line is for
the weakly excited BEC (b). The collisional products of the
strongly driven BEC are clearly driven towards the center, while
those of the weakly driven BEC are concentrated on the s-wave
sphere ($y=x_r$).}
\end{figure}

In our experiment, however, the condensate is \emph{strongly}
driven at large momentum $2k_L\xi=3.9$, and collisions occur
within the lattice potential. Consequently, the Bogoliubov
description is no longer adequate. This is clearly visible in Fig.
\ref{fig:tomog}(a), which shows an absorption image obtained after
a resonant dressing pulse lasting $t=660$ $\mu$s. Upon comparison
with the s-wave sphere obtained when two condensates collide [Fig.
\ref{fig:tomog}(b)], one sees  a clear shift of the collided atoms
towards the center of the sphere. To quantify this difference we
employ computerized tomography to extract the radial dependence of
the density from the column-density available in the absorption
image \cite{ours-tomography}. In Fig. \ref{fig:tomog}(c) we plot
the radial distribution of atoms over a small slice in $\hat{z}$.
This inward shift of collided atoms is robust and is clearly
observed for different values of $\Omega_d$ and $t$.

\begin{figure}[h]
\begin{center}
\includegraphics[width=7cm]{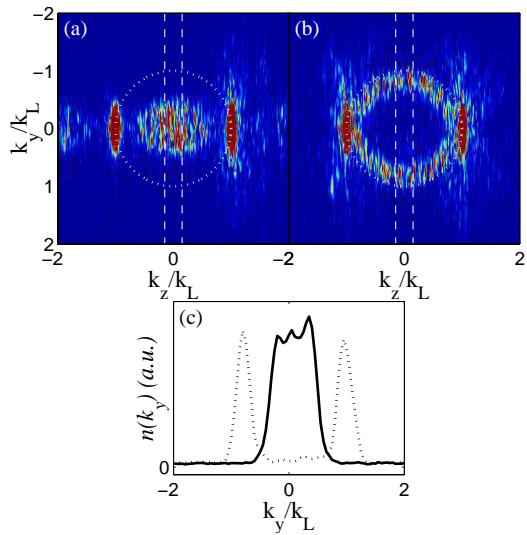}
\end{center}
\caption{ \label{fig:stochastic} (a,b) Momentum distributions
generated by the classical field GPE simulation (a) Strong pulse
$\Omega_d=8.6$ kHz, pulse width $t=660$ $\mu$s (b) Weak pulse
$\Omega_d<2$ kHz, pulse width $t=370$ $\mu$s. The strong lattice
leads to a clear shift of the collisional products inward as
compared to the weak pulse, and deviates strongly from the s-wave
collisional sphere. Note that the simulation also generates
momentum wavepackets with clear number correlations. (c) The
density distribution along the $y$-axis, of the simulation,
averaged over a slice marked by the vertical dashed lines. The
solid line is for the strongly driven BEC of (a), and the dotted
line is for the weakly excited BEC (b). The inward shift, observed
in the experiment [Fig. \ref{fig:tomog}(c)], of the decay products
is even more pronounced here.}
\end{figure}

We qualitatively simulate this collisional decay using the
stochastically seeded classical field method, in 2D, as it was
developed recently for colliding BECs \cite{gardiner}. In this
method the initial seed of fluctuating random amplitudes of the
bosonic field is added to the ground state of the condensate in
the harmonic trap. Then the moving lattice potential is switched
on suddenly (as in the experiment). Matter wave mixing between the
condensate momentum wavepackets and the seeded quasi-continuum
drives the collisional decay of the oscillations, without any need
for further gross numerical intervention. Fig.
\ref{fig:stochastic}(a) shows the resulting momentum distributions
for collisions occurring within the lattice, and Fig.
\ref{fig:stochastic}(b) shows the collisional manifold when only a
weak lattice is present. We note that the overall agreement
between the simulation and experiment takes into account many
possible systematic effects such as the harmonic confinement in
the radial dimension and finite time of the Bragg pulse. We also
see that the essential physics is qualitatively captured here,
even though the simulation is not in 3D. The correlations between
counter-propagating momentum wavepackets, clearly visible in the
simulation, are not visible in the experiment due to the fact that
the experimental absorption images integrate over an additional
dimension, making this signal difficult to observe. The mean-field
broadening of the experimental time-of-flight images also leads to
some additional smearing.

To obtain an intuitive model that still captures the essence of
these phenomena, we neglect inhomogeneous and finite-size effects
and choose our frame of reference as moving with the optical
lattice of the Bragg lattice beams at velocity $v=-\hbar k_L/M$
along $\hat{z}$. Our system is then described by the many-body
time-independent Hamiltonian

\begin{eqnarray}
H=&\sum_\textbf{k}\left[\frac{\hbar^2
k^2}{2M}\creation{a}_\textbf{k}\hat{a_\textbf{k}}
+\frac{\hbar\Omega_d}{2}\left(\creation{a}_\textbf{k}\hat{a}_{\textbf{k}-2\textbf{k}_L}+\creation{a}_\textbf{k}\hat{a}_{\textbf{k}+2\textbf{k}_L}\right)\right]
\nonumber \\
&+\frac{g}{2V}\sum_{\textbf{k},\textbf{l},\textbf{m}}\hat{a}_\textbf{k}^{\dag}\hat{a}_\textbf{l}^{\dag}\hat{a}_\textbf{m}\hat{a}_{\textbf{k}+\textbf{l}-\textbf{m}},\label{eq:many-body}
\end{eqnarray}
where $\creation{a}_\textbf{k} (\hat{a_\textbf{k}})$ is the
creation (annihilation) operator of a particle with wave-vector
$\textbf{k}$, and $g$ is a constant describing the s-wave
interactions. The relative momentum in the experiment is
sufficiently low to avoid higher partial wave collisional terms
\cite{wilson,walraven}. Neglecting for the moment the interaction
term, Eq. (\ref{eq:many-body}) simplifies into a one-dimensional
single particle Hamiltonian,

\begin{equation} H=-\frac{\hbar^2}{2M}\frac{\partial^2}{\partial z^2}+\hbar\Omega_d
\cos(2k_Lz) \label{eq:bloch hamiltonian}.
\end{equation}

According to Bloch's theorem, a state $\ket{k}_w$ is only coupled
to states $\ket{k+2pk_L}_w$, where $p$ is an integer.  In the
moving frame of reference, the stationary BEC has momentum $\hbar
k_L$ and is situated on the Brillouin zone boundary. The initial
kinetic energy of the condensate is therefore in the lattice
energy gap. To calculate the consequent dynamics we span the state
$\ket{k_L}_w$ by the new basis of Bloch states $\ket{n}_b=\sum_p
a_{n}((2p+1)k_L)\ket{(2p+1)k_L}_w$, which diagonalize the
Hamiltonian (\ref{eq:bloch hamiltonian}) \cite{phillips-lattice},
where $n$ is the Bloch band index. The lattice momentum $\hbar
q=\hbar k_L$ remains unchanged and is therefore omitted. The
subscript $w\text{ and }b$ indicate whether the quantum numbers in
the ket describe the wavenumber of a plane-wave or a Bloch band
index.

In the weak lattice limit, we arrive at the two state result
$|k_L\rangle_w=1/\sqrt{2}\left(\left|1\right\rangle_b+\left|2\right\rangle_b\right)$,
of a two level system undergoing Rabi oscillations with frequency
$\Omega_d$. This two mode picture is still useful even for
stronger lattices, since the energy separation between the lower
two bands to the third band, on the edge of the Brillouin zone, is
typically larger than the $\Omega_d$'s discussed here. Therefore,
the higher bands are only weakly occupied by the system.

We now consider the mixing of Bloch states due to atomic
interactions. In order to describe collisions, we include the
interaction term in the Hamiltonian (\ref{eq:many-body}), which
scatters two atoms from the populated states. We focus on the
processes in which atoms are scattered into the quasi-continuum of
unpopulated states, neglecting scattering into populated states
(forward scattering) \cite{neglect}. Since collisions are binary
the Hilbert space is reduced to a two particle space. Hamiltonian
(\ref{eq:many-body}) can be rewritten in the basis of the Bloch
Hamiltonian (\ref{eq:bloch hamiltonian}) as
\begin{eqnarray}
H=\sum_{\nu_1,\nu_2}\ket{\nu_1;\nu_2}\bra{\nu_1;\nu_2}\left[E_{\nu_1}+E_{\nu_2}\right]+\nonumber\\
\frac{g}{2V}\!\!\!\!\!\!\!\!\sum_{\textbf{k}_1\ldots\textbf{k}_4,\nu_1\ldots\nu_4}\!\!\!\!\!\!\!\!
\braket{\nu_1}{\textbf{k}_1}\braket{\nu_2}{\textbf{k}_2}\braket{\textbf{k}_3}{\nu_3}\braket{\textbf{k}_4}{\nu_4}\times
\nonumber \\
\ket{\nu_1,\nu_2} \bra{\nu_3,\nu_4}
\delta_{(\textbf{k}_1+\textbf{k}_2)-(\textbf{k}_3+\textbf{k}_4)}.\label{eq:scatteringH}
\end{eqnarray}
Here $\nu_i$ stands for all quantum numbers of a Bloch state
$n_i,q_i,\textbf{k}_{i\perp}$. $E_{\nu_i}=E_{n,q}+\left(\hbar
k_{i\perp}\right)^2/2M$ is the energy of the noninteracting Bloch
state, where $E_{n,q}$ is an eigenvalue of the Bloch Hamiltonian
(\ref{eq:bloch hamiltonian}) and
 $\textbf{k}_{i\perp}$ is the part of $\textbf{k}_i$ which is
perpendicular to $\hat{z}$ \cite{bosamp}. The inclusion of the
quantum numbers $\textbf{k}_\perp$ is necessary since collided
atoms gain momentum which is not along $\hat{z}$. We treat the
collision term in Eq. (\ref{eq:scatteringH}) by use of
perturbation theory \cite{secondary}. That is, we assume the
system is undergoing coherent oscillations in time due to the
lattice potential, and study the perturbative collisional products
that are created by the interaction Hamiltonian.

The existence of two macroscopically occupied, distinct energy
states, implies several decay routes for the collisional term, and
the subsequent energy and momentum conservation manifold are
split. Specifically, splitting arises from the energy difference
between the case where both colliding atoms are from the $n=1$
band and when both are from the $n=2$ band. Due to symmetry,
destructive interference suppresses collisions in which there is
initially one atom in each of the two bands.

\begin{figure}[h]
\begin{center}
\includegraphics[width=7cm]{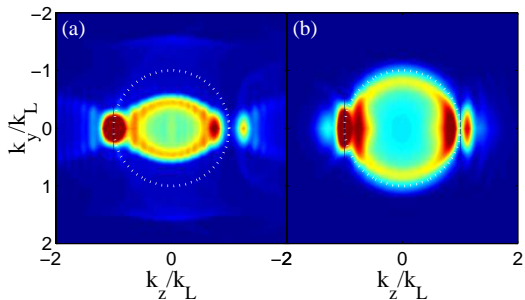}
\end{center}
\caption{ \label{fig:model} Column momentum distribution for the
same parameters as for Fig. \ref{fig:tomog}(a) and
\ref{fig:tomog}(b) respectively calculated by our colliding Bloch
state model (Eq. \ref{eq:scatteringH}, only the scattered atoms
are shown). Dotted line represents the s-wave shell. The momentum
distribution is only roughly equivalent to the spatial
distributions of Fig. \ref{fig:tomog} due to interactions during
expansion, although the overall shape of the collisional manifold
is reproduced.}
\end{figure}

The prediction of our model for the momentum distribution of the
collisional products is plotted in Fig. \ref{fig:model}.  The
calculated column density is presented for comparison with
Fig.~\ref{fig:tomog}.

The presence of the optical lattice is found to drive the collided
atoms towards the center, as expected from the inner shell of the
splitting. The amplitude of the outer shell decreases rapidly as
$\Omega_d$ increases, and is experimentally unobservable for our
parameters. In the experimental data additional effects such as
inhomogeneous broadening and mean-field expansion effects, that
are neglected in this model, tend to broaden the shell of atoms,
and blot out the splitting for smaller $\Omega_d$.

One intriguing prediction of the model, is that the decay rate as
a function of time will deviate significantly from that predicted
by the Fermi golden rule. This can be explained in the time domain
by the oscillatory behavior of the coherent evolution, leading to
a complementary oscillation in the rate of production of collided
atoms.

Another important result of the model is that as we increase
$\Omega_d$, the overall decay of the coherent evolution is
accelerated. This is due to the large number of additional decay
pathways that are switched on by the presence of a deeper lattice.

In conclusion, we measure the collisional decay of a driven
condensate and show that it deviates from the usual s-wave sphere.
This result is modelled by using the stochastically seeded
classical field method applied to the Gross-Pitaevskii simulation
of the experiment. The main features of the collisional decay
manifold are captured by a simple model, which treats the
interactions as binary collisions between single particle Bloch
states.

This work was supported in part by the Israel Ministry of Science,
the Israel Science Foundation, and the DIP Foundation. We thank C.
Gardiner and A. Norrie for useful discussions.


\begin{thebibliography}{99}
\bibitem{neglect} Although the weight of such processes is of order $N$ larger than
  the former, this approach is justified by the observation that the meanfield shift
  is being averaged out by the rapid oscillations \cite{ours-dephasing}. Its contribution is therefore suppresed.
\bibitem{bosamp} In general each interaction term in Eq. (\ref{eq:scatteringH})
  should be multiplied by an appropriate bosonic amplification factor. However, in our
  system there are two approximately equally populated initial Bloch
  states and target states are assumed to be empty, therefore all terms in
  Eq. (\ref{eq:scatteringH}) should be multiplied by the same $\sim N/2$.
\bibitem{secondary} Using a 3D simulation of the classical collisional and expansion
  dynamics, we estimate that less than 10 percent of the atoms in
  Fig.~\ref{fig:tomog}(c) underwent secondary collisions, and therefore the
  distribution is dominated by single collisions as discussed in our model.
\end{thebibliography}
\end{document}